\documentclass[10pt]{article}
\usepackage{epsfig}
\usepackage{aas2pp4}
\usepackage{aassize}

\newcommand{\cf}{{\it cf.}}
\newcommand{\eg}{{\it eg.}}
\newcommand{\et}{{\it et~al.}}

\newcommand{\CGRO}{{\it CGRO}}
\newcommand{\EGRET}{{\it EGRET}}
\newcommand{\COSB}{{\it COS~B}}

\newcommand{\text}[1]{{\rm #1}}
\newcommand{\eff}{$\eta_\gamma$}
\newcommand{\beam}{$f_\gamma$}

\newcommand{\Gam}{$\gamma$ rays}
\newcommand{\gam}{$\gamma$-ray}

\begin{document}

\title{$\gamma$-Ray Pulsars and Massive Stars in the Solar Neighborhood}
\vspace{-0.4cm}
\author{I.-A. Yadigaroglu and Roger W. Romani\altaffilmark{1}}
\affil{Department of Physics, Stanford University, Stanford, CA 94305-4060}
\altaffiltext{1}{Alfred P. Sloan Fellow}

\vspace{-.5cm}
\begin{abstract}
We revisit the association of unidentified Galactic plane \EGRET\
sources with tracers of recent massive star formation and death.
Up-to-date catalogs of OB associations, SNRs, young pulsars, HII
regions and young open clusters were used in finding counterparts for
a recent list of \EGRET\ sources. It has been argued for some time
that \EGRET\ source positions are correlated with SNRs and OB
associations as a class; we extend such analyses by finding additional
counterparts and assessing the probability of individual source
identifications.  Among the several scenarios relating \EGRET\ sources
to massive stars, we focus on young neutron stars as the origin
of the \gam\ emission. The characteristics of the candidate
identifications are compared to the known \gam\ pulsar sample and
to detailed Galactic population syntheses using our outer gap pulsar
model of \gam\ emission. Both the spatial distribution and luminosity
function of the candidates are in good agreement with the model
predictions; we infer that young pulsars can account for the bulk
of the excess low latitude \EGRET\ sources. We show that with this
identification, the \gam\ point sources provide an important new
window into the history of recent massive star death in the solar
neighborhood.
\end{abstract}

\vspace{-0.2cm}
\vspace{0.1cm}

\twocolumn

\section{Introduction}

At least two distinct populations of sources have been detected above
100 MeV by the \EGRET\ experiment on \CGRO: extragalactic AGN, and a
population of $\sim$~30 Galactic sources with small scale height. Five
of the Galactic sources have now been identified from their pulsations
as newly born neutron stars. It has been known for some time (Lamb
1978, Montmerle 1979) that the Galactic population is, {\em as a
class}, closely linked with young objects. Montmerle showed that about
half of the 11 unidentified Galactic \COSB\ sources could be
associated with ``SNOBs,'' spatial coincidences between SNRs and OB
associations (or HII regions as their tracers). Conversely, as many as
three-fourths of the best identified SNOBs were seen in \Gam. However,
SNRs as a class were clearly not all \gam\ emitters. In principle
\Gam\ from the SNOBs could be due to any combination of features of
young regions, and the particular scenario advanced by Montmerle was
that the dense neighborhoods of massive star associations act as
targets for the cosmic rays produced in shock waves of SNRs, resulting
in \Gam\ from $\pi^0$ decay.

Surprisingly, even though two of the brightest sources had already
been identified with the Crab and Vela pulsars, it was not proposed
that the other sources are also young

\newpage
\mbox{}
\begin{minipage}[l]{3in}
\vspace{5.1in}
\end{minipage}

\hspace{-0.43cm}isolated pulsars. Since then,
the \gam\ source positions have of course been extensively searched
for radio pulsations, and, conversely, the known young radio pulsars
have been searched for \Gam\ (see Thompson 1994 for recent upper
limits). Only three new pulsars with both radio and \gam\ emission
have been found in this manner (PSRs B1706-44, B1055-52 and
B1951+32). Even though it is notoriously difficult to detect short
period radio pulsars in the dense interstellar medium of massive star
forming regions, the small number of identifications would seem to
cast doubt on pulsars as the origin of all Galactic \gam\
sources. However, the discovery of the Geminga \gam\ pulsar from its
X-ray pulsations (Halpern and Holt 1992) has dramatically altered our
picture of the \gam\ pulsar population, as it presents no observed
radio emission. Several authors have thus recently revisited the
question of whether all the unidentified \gam\ sources are in fact
isolated neutron stars waiting to be discovered as such.

Duplication of the discovery in X-rays of the Geminga pulsar to other
sources has proven difficult, and direct searches of pulsations in
\Gam\ have only recently begun to place limits on the pulsed fraction
and spin parameters of the brightest few sources (Mattox \et\
1996). Until more high energy data becomes available population
studies are thus an interesting alternative in exploring the nature of
the unidentified sources.

Halpern and Ruderman (1993) and Helfand (1994) assume simple forms for
the evolution with age of the efficiency \eff\ of conversion of
spin-down power to \Gam\ and single values for the beaming fraction
\beam\ of \Gam\ on the sky and conclude that all \gam\ sources may well
be pulsars. Mukherjee \et\ (1995) arrive at the opposite conclusion by
estimating the distance and luminosity of the unidentified \EGRET\
sources from their distribution in Galactic longitude and
latitude. Assuming again a unique value for \beam, they find that
luminosities of the unidentified sources are too large to be
Geminga-like objects. In Romani and Yadigaroglu (1995, RY) we
presented an outer gap model for the emission of high energy \Gam\ by
pulsars younger than a million years. In our model both \eff\ and
\beam\ evolve with pulsar age, and the beaming pattern differs from
that of the radio emission. Monte Carlo sums (Yadigaroglu and Romani
1995, YR) showed that most of the unidentified \EGRET\ sources are
expected to be Geminga-like pulsars with no detectable radio emission.
Other pulsar \gam\ emission models, such as the extended polar cap
models of Daugherty and Harding (1996) and Sturner and Dermer (1996),
predict that at most a few of the unidentified sources are pulsars,
many of which should have faint radio emission.

Kaaret and Cottam (1996, KC96) have recently revisited Montmerle's
hypothesis. They find that 16 of 25 unidentified \EGRET\ sources lie
in or near OB associations, with a probability that the superposition
as a class is due to chance of $10^{-4}$. Several of the 16 sources
are also noted to have coincident SNRs and/or radio pulsars.  The
authors estimate distances to the \EGRET\ sources from the association
distances, and construct an intrinsic luminosity function. The
luminosity function is found to be consistent with the known
\gam\ pulsars, leading the authors to conclude that a majority
of the \EGRET\ unidentified sources are probably pulsars.

Sturner and Dermer (1995, SD95) have searched for coincidences between
unidentified \gam\ sources and SNRs using a much improved test of
association. Along with Esposito \et\ (1994) they find a statistically
significant correlation.  All of the SNRs found to be coincident with
\EGRET\ sources are in fact SNOBs, and are present in the association
lists of both Montmerle and KC96.  The view adopted in SD95 is that
the \gam\ emission originates from the remnants and three possible
mechanisms are suggested: pulsar powered plerionic emission from
filled-center SNRs, and for shell-type SNRs either non-thermal
bremsstrahlung (from the synchrotron emitting electrons) or $\pi^0$
decay from remnant-generated cosmic rays colliding with the
interstellar medium. Note that as in the SNOB hypothesis, the
\gam\ sources should then be extended. To date this has only been
tested for 2EG J2020+4026 (Brazier \et\ 1996) which is unresolved,
in contrast to the associated SNR $\gamma$ Cygni.

We update these studies with an improved version of the test statistic
of SD95 (\S\ref{Test}), finding coincidences between the most recent
\EGRET\ source list and an up-to-date catalog of young region tracers,
including OB associations, SNRs, young pulsars, HII regions, and young
open clusters (\S\ref{Catalog}). We present combined results for all types
of objects, and construct a luminosity function for the sources from
the estimated distances to the counterparts (\S\ref{Candidate}). We then
focus on young pulsars, and compare a Monte Carlo population synthesis
(\S\ref{Simulation}) using our outer gap model and a detailed model of
the solar neighborhood with the luminosity function and other
characteristics of the now ``identified'' sources (\S\ref{Compare}).

\section{\label{Test}Test of Association}

We have extended Sturner and Dermer's test of association between
\EGRET\ sources and SNRs to allow a general search for counterparts.
Their statistic is $\alpha_1 = (r_1/r_2)^2$, where $r_1$ and $r_2$ are
the angular distances from an unidentified source to the nearest and
second nearest young object. Interpreting $2/r_2^2$ as an estimate of
the local density of young objects $n_y$, $\alpha_1$ will then be
distributed uniformly between 0 and 1 for random populations, and
small values of $\alpha_1$ will indicate unlikely coincidences. The
``$1\;\sigma$ confidence interval'' on $\alpha_1$ is simply
$\sigma_{\alpha_1}~=~\pm~0.32$.

Sturner and Dermer's test has the advantage that it does not require
an independent estimate of the local density of a given object class,
as obtained for instance from CO maps (\cf\ KC96). Such estimates do
not eliminate unknown observational biases intrinsic to the object
catalogs. While keeping this useful property we have taken into
account the angular size of the objects as well as errors in the
determination of the source position.

We use the \EGRET\ source location 2D Gaussian probability
distribution and construct its overlap integral Q with a second 2D
Gaussian distribution associated with each young object in our
catalogs. For counterparts other than pulsars, the width of this
second Gaussian is set to the object's angular size, $s$. The integral
Q is thus the probability that the \gam\ source position is within
the boundary of the object, with the object position weighted as a
Gaussian about its nominal center.

The overlap integral is then normalized by the expected number of
catalog sources within the \EGRET\ error circle and the resulting
ratio compared with the distribution of ratios expected for a random
population of similar young objects at density $n_y$. As in SD95,
$n_y$ is estimated from nearest neighbors in the catalog (we have
chosen to use the third nearest neighbor, since several catalog
sources are typically within the scale of Galactic structures). So for
each unidentified \EGRET\ source, we find the nearest young object in
a counterpart catalog, and report the probability P that a larger
ratio would have been found by chance (P[Q,$\,n_y$] was obtained from
Monte Carlo sums). For uncorrelated samples these probabilities are
uniformly distributed, and with 35 \EGRET\ sources, we should find one
pair with P~$<$~3~\%. In fact since $\sim$~10 sources have very small
P, less than one pair is expected to have P~$<$~3~\% due to chance.

\section{\label{Catalog}Object Catalogs}

\subsection{ \EGRET\ sources }

The Second \EGRET\ Catalog (Thompson \et\ 1996) lists 71 unidentified
\gam\ sources, 33 of which fall within $|b|~<~10^\circ$ of the
Galactic plane.  In finding associations, we have used the updated
list of Thompson \et\ 1996b. This list benefits from additional phase
III exposure, and lists 8 additional sources with $|b|~<~10^\circ$.
In our analysis we have eliminated sources for which the {\em average}
flux in the three phases results in a significance $\sqrt{\text{TS}}$
of less than 5 (approximately $5\;\sigma$, $\sqrt{\text{TS}}$ as defined
in Thompson \et\ 1996), as this is the threshold used in modeling the
\EGRET\ sensitivity in section \S\ref{Simulation}.  The error ellipses
for the source positions are given in Thompson \et\, and their mean radius
is $\sim~30'$. Uncertainties remain in understanding \EGRET\ likelihood
distributions, and a careful calculation using the detailed shape of
the error contours instead of 2D Gaussians will clearly be of benefit
to an update of the present counterpart search.

\subsection{ OB Associations and Young Clusters }

For a catalog of OB associations, we follow KC96 in using the list of
Mel'nik and Efremov (1995).  The O and B-stars of Blaha and Humphreys
(1989) are partitioned by Mel'nik and Efremov into 58 OB associations
using a cluster analysis method. Of the 630 O-stars present in the
star catalog, $\sim$~75~\% are found to lie in associations. For each
OB association, the mean position, distance, size, and number of stars
are given. The associations have a mean size of 40 pc and a mean
distance of $\sim$~1.6~kpc. Associations with size less than $\sim$~15
pc are defined as young open clusters.  We note that distances to O
and B-stars are accurate to only $\sim$~20~\%.

It is well-known that large OB associations are complex and can often
be partitioned further into sub-structures, see Garmany (1994) for a
review.  The algorithm of Mel'nik and Efremov finds these
sub-structures and reports them as separate OB associations. In their
nomenclature, these fragments will have the same group name with the
added suffices A, B, C, etc. As we estimate the local density of OB
associations for our test statistic from the distance to the third
nearest unassociated neighbor, we ignore fragments from the same
association in selecting nearest neighbors.

In addition to the four open clusters found by Mel'nik and Efremov we
have used the 160 open clusters younger than $10^{7.5}$~years in the
catalog of Janes, Duke and Lyng\aa\ (1987). The mean distance to the open
clusters is $\sim$~2.1~kpc.

\subsection{ \label{HIIreg}HII regions }

HII regions and their associated molecular clouds are convenient
additional tracers of O and B-stars. The observational situation has
not changed much since Montmerle's (1979) study, and we have also used
Georgelin and Georgelin's (1976) catalog of 100 giant HII regions.
Georgelin and Georgelin map the spiral structure of the Galaxy and
thus select HII regions that would be prominent to an observer external
to our Galaxy, so that their sample includes only intrinsically large
and bright HII regions.  Distance and class are reported in the tables
of Georgelin and Georgelin.  Region class reflects the radio size and
brightness of the region and is one of `f', `m' or `b'. For 40 of the
50 HII regions within 5~kpc, the region size could be determined from
Mar\u{s}\'{a}lkov\u{a} (1974) and other original sources. For the distant
radio selected regions, we have assumed a diameter of 50 pc. Many of the
giant HII regions are complex associations of smaller HII regions.

Brand and Blitz (1993) have compiled a much more detailed but local
sample of 206 kinematically distinct HII regions from various
optically selected catalogs. The HII region sizes are not given
in this catalog but were available from the original references in
most cases.  The mean distance of the local sample is $\sim$~2.6~kpc.
We do not expect to find good (small P) \EGRET\ counterparts in the
local sample of HII regions since it distinguishes between different
sub-structures of the same large massive star forming region, skewing
the test estimate of the local density of unassociated HII regions.

\subsection{ Young Radio Pulsars and SNRs }

O and B-stars are the direct progenitors of neutron stars, and five
\EGRET\ sources are presently identified with radio pulsars: the Crab
pulsar, Vela, PSRs B1706-44, B1055-52 and B1951+32, all younger than
$10^6$~years. Extensive \gam\ searches of all known radio pulsars have
of course been performed (Thompson \et\ 1994, Fierro 1996).  For
completeness and as a test of our statistic we have looked for
associations between unidentified sources and young radio pulsars. An
extensive catalog of radio pulsars is available from the Princeton
Pulsar Group (see Taylor, Manchester and Lyne 1993) and contained 104
pulsars younger than $10^6$~years at the time of our analysis.

The Princeton database is collated from many different radio surveys
and thus has non-uniform flux limit. The pulsar sample has a mean
distance of $\sim$~6~kpc but is certainly not complete to this
distance.  There are strong selection effects for very young radio
pulsars that bias the sample towards long periods and high magnetic
fields. In addition radio pulsars are difficult to detect in regions
of high electron density $n_e$, such as in complexes of SNRs, HII
regions and O and B-stars.

SNRs generally persist for $\lesssim\;10^5$~years and are ideal tracers
of recent massive star death. We use the latest SNR catalog of Green
(1995) containing 194 SNRs. SNR position and angular size were taken
from the catalog, but we adopt distance determinations from HI
absorption or Sedov solutions in the literature, when available. When
no estimate was available, we compute the $\sigma-D$ distance which is
notoriously inaccurate. In the near future more accurate Sedov X-ray
distances should be available from the ROSAT all-sky survey, see Namir
\et\ (1994). The SNR sample is collected from many sources and has a
mean distance of $\sim$~3~kpc. Many additional SNRs are expected from
the ROSAT survey, which will result in a deeper sample.

\section{\label{Candidate}Coincident Sources}

The complete list of \gam\ sources and their candidate associations
are summarized in Table~1.  For each \EGRET\ source, we list the
source position, flux, and all catalog objects that are consistent
with the source position, rejecting pairs for which Q~$<$~10~\%
(i.e. for the given source size and \EGRET\ error circle, association
can be rejected at the 90~\% level).  The probability P (defined in
\S\ref{Test}) that the coincidence is due to chance is listed for each
pair. We have high confidence in the identification of the \EGRET\
source if at least one probability P is $<$~3~\%. We distinguish
between these ``IDed'' sources (bold type) and ``overlaps'' for which
the P of all coincident objects are $>$~3~\%. Unfortunately, when
several young objects are clustered in the same region of the sky, as
often occurs along tangents to the spiral arms, we cannot ID an
\EGRET\ source with a unique object even though the source is probably
associated with the region as a whole. Some overlaps are thus likely
to be real coincidences. When P~$>$~10~\% we do not list P, but
only Q.

\begin{figure}[htb]
\plotone{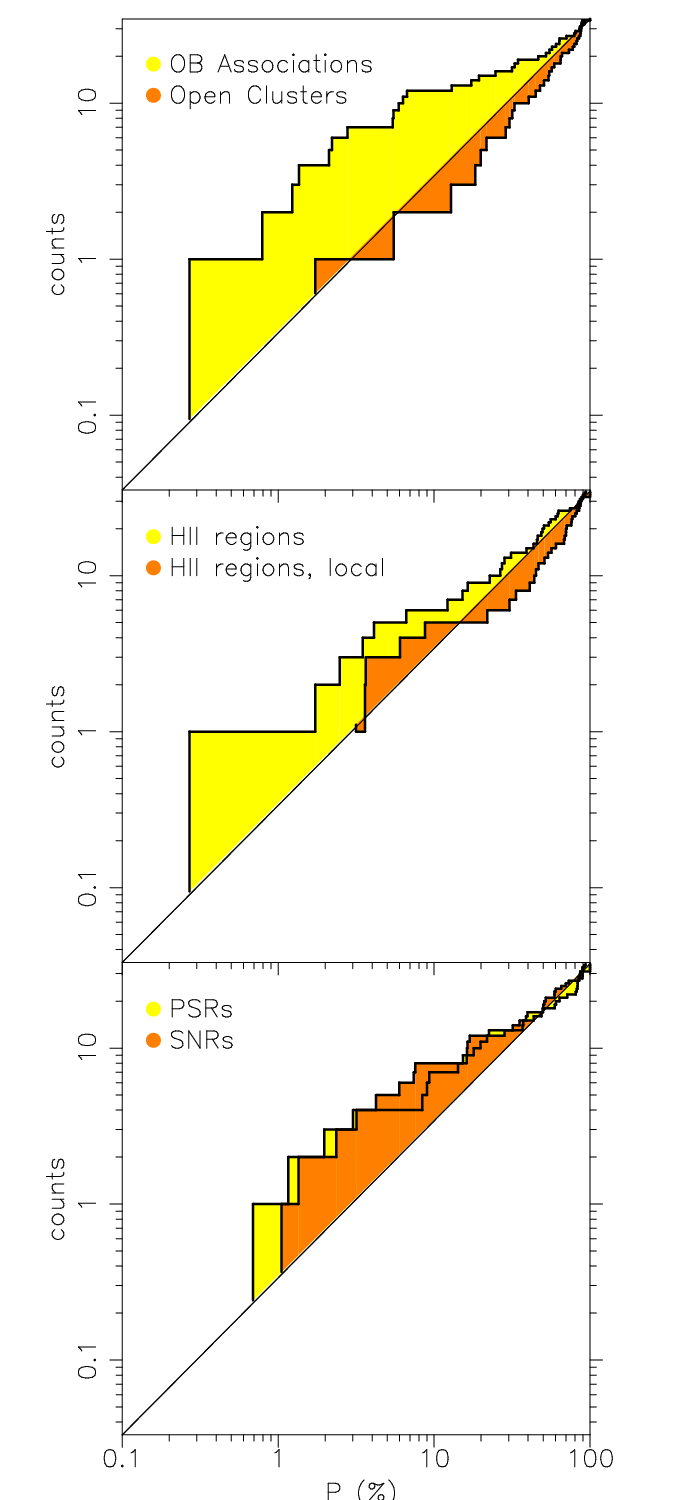}
\caption{Cumulative distributions of association probabilities. The
value P is the probability that an association pair is due to
coincidence assuming uniformly distributed objects locally. The total
number of unidentified sources is 35, so at each P we can expect $35
~\times$~P objects to have this value of P by chance if the pairs are
unassociated; this value is denoted by the diagonal line. The
shaded surface thus represents the excess number of associations.}
\label{cumulative}
\end{figure}

In Figure~\ref{cumulative} we plot the cumulative distributions of the
probabilities P of association. For uncorrelated samples these
probabilities should be uniformly distributed and would follow the
diagonal line. Clearly all but the young open cluster and local HII
region samples present significant numbers of improbably close
counterparts. The largest excess of counterparts is found for the OB
associations, followed by SNRs. Thirteen \EGRET\ sources have at least
one P~$<$~3~\%, and 20 have P~$<$~10~\%. Of the remaining sources, 3
have overlapping objects with Q~$>$~10~\%, so that only 12 \EGRET\
sources have no coincident objects at all in the examined catalogs.

Our test has of course recovered the five pulsed identifications with
known pulsars with very high confidence (P~$<$~0.01~\%); PSR~B1951+32
is too faint to be included in the catalog as an unpulsed source. These
are not included in Figure~\ref{cumulative} or Table~1. The Crab Nebula,
Vela and PSR~B1706-44 remnants (MSH 17-41) are also found with high
confidence (P~$<$~3~\%). Only the Vela pulsar has an OB association
(OB Vela 1A) consistent with its position. No other coincidences are
found. This is not surprising given the difficulty in detecting radio
pulsations from regions of high electron density $n_e$.

Comparing with previous results, we have found all 7 SNR counterparts
reported in SD95 for \EGRET\ sources still present in our source
list. However, only three of the seven, SNRs IC 443, MSH 11-61A and
$\gamma$ Cygni result in source IDs, and the coincidence with SNR
G312.4-0.4 is just above our ID cut of P~$<$~3~\%. We have also found
all of the 15 OB counterparts reported in KC96 for \EGRET\ sources
in our list. Six of the 15 result in IDs (Gem 1, Car 1 F, Sgr 1 C,
Sgr 1 B, and Cyg 1,8,9 twice), and 10 have P~$<$~10~\% (add Mon 1 B,
Clust 3, Car 1 A, and Cyg 1,8,9 again). Seven of the OB counterparts
are also coincident with a SNR.

For many of these \EGRET\ sources, we have also found nearby HII
regions. In addition we have seven new \EGRET\ IDs or overlaps. Two
of these are \EGRET\ sources not present in earlier catalogs. No
sources are IDed with an HII region from the local sample, as expected
(see \S\ref{HIIreg}). Only one source is IDed with a young open cluster,
Collinder 347. This indicates that neutron star formation ceased over
a million years ago in most of the open clusters in our catalog. 

\begin{figure}[t]
\plotone{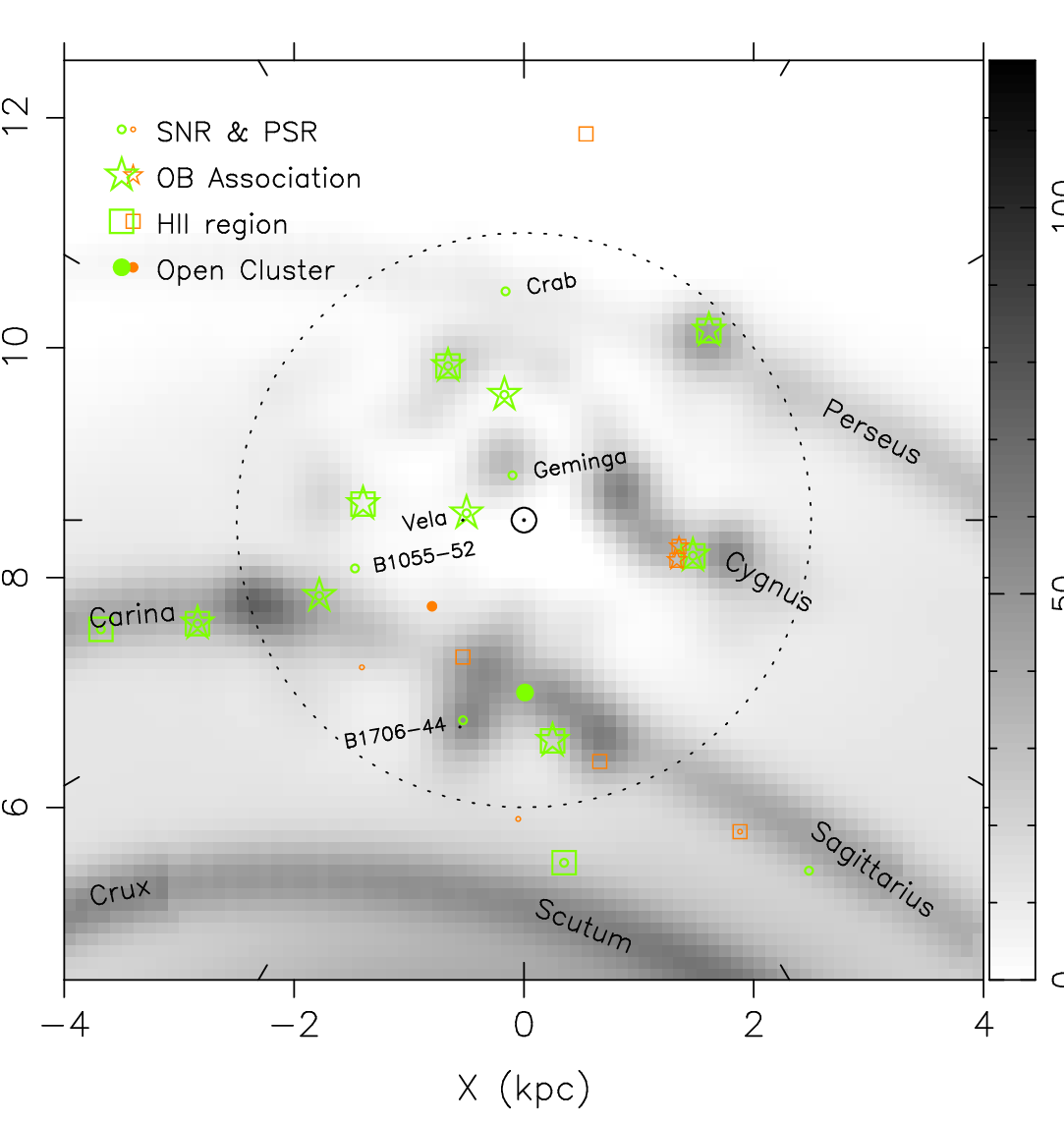}
\caption{Distribution of candidates in the Galactic plane, viewed from
directly above the Sun. Larger symbols are for young objects
associated with an unidentified \EGRET\ source with high probability
(IDs), and smaller symbols for overlap candidates.  The background
gives the modeled distribution of O-stars.  Shading is normalized to
show the corresponding number of neutron stars per kpc$^2$ younger
than $10^6$~years (for a birth rate of $1/100$~years). The radial tick
marks are for Galactic $l$ at $30^\circ$ intervals, counter-clockwise
from the bottom.}
\label{above}
\end{figure}

Having found the list of \gam\ sources with candidate associations, we
have assigned a distance $d$ to each of the 23 IDs $+$ overlaps.  If a
counterpart with P~$<$~10~\% was available, we chose the distance to
the counterpart with smallest P.  Otherwise we chose the distance to
the counterpart with largest Q. The Galactic positions of these 22
sources along with the five known \gam\ pulsars are plotted in
Figure~\ref{above}. Note that when there are several counterparts for
the same \EGRET\ source it is often possible to identify a single
distance with the entire complex.

We can now calculate the \gam\ luminosities for our candidate sources
from the measured photon fluxes and estimated distances. The source
spectra are often not well fit by a power law, and rather than assume
a spectral index we have numerically summed the spectra above 100 MeV
as given in Merck \et\ (1996) (two of our IDed sources are not present
in Merck's list and have not been included in the constructed
luminosity function). An upper cutoff of 3~GeV was assumed when only
upper limits on the high energy source flux were available.  We have
also assumed isotropic radiation (\beam~=~1). Typical isotropic
luminosities are several $\times$~$10^{35}$~erg/s, similar luminosities
have recently been inferred by KC96 and Kanbach \et\ (1996). For pulsars
in reality we expect \beam\ as small as 0.1 in many cases (YR, R96);
``isotropic'' luminosities can therefore exceed the total spin-down
power.

We note that pulsars would be expected to have indices of -1.7, but
many may have an additional soft component due to plerionic emission
or unresolved diffuse excesses from gas clumps or associated cosmic
ray enhancement that will steepen the measured index. It is thus not
surprising that Merck \et\ find only a few spectral indices to show
the characteristic hardness expected for pulsars. However, it is
interesting that the handful of sources with hard spectra reported on
in Merck \et\ are independently singled out in our analysis as being
good (low P) candidates.

\begin{figure}[t]
\plotone{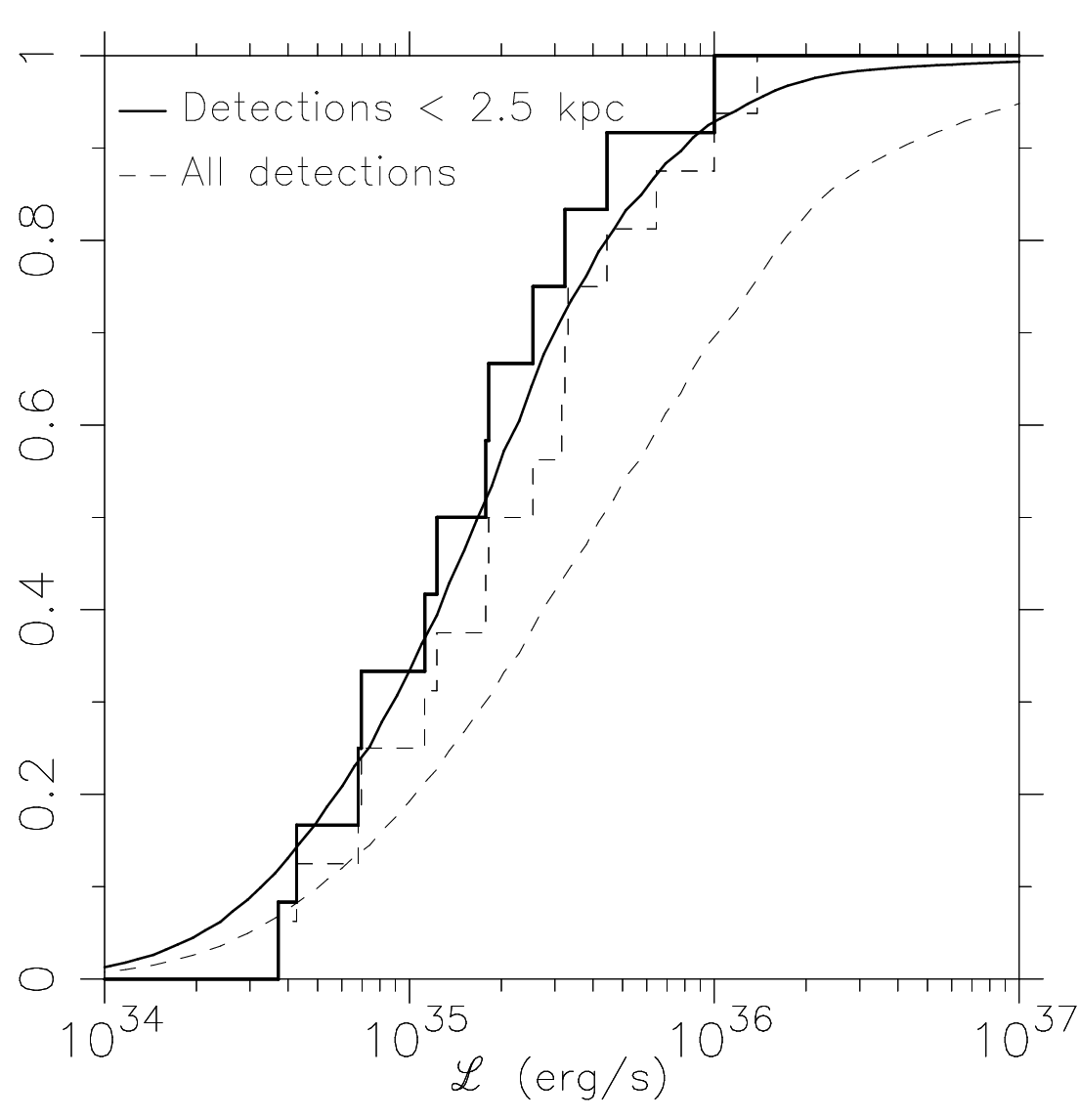}
\caption{Luminosity functions for the IDed and modeled populations.
Both complete and truncated ($<$~2.5~kpc) luminosity functions are
shown.  Only the truncated luminosity functions should be compared as
the IDs are incomplete beyond 2.5~kpc.}
\label{luminosity}
\end{figure}

\section{\label{Simulation}Modeled Pulsar Population}

We now turn to spin-down powered pulsars as the origin of the \gam\
emission.  For the evolution of efficiency of spin-down power to \Gam\
we assumed in RY and YR a simple empirical law that matches the values
for PSRs B1706-44 and B1055-52. A recently completed model of emission
processes in the outer magnetosphere (Romani 1996, R96) gives a more
complete description of the pulsar radiation. In R96 the efficiency of
curvature radiation \eff\ in the \EGRET\ range is estimated as a
function of age, magnetic field and inclination angle.  The \eff\ and
\beam\ laws of R96 thus allow a more accurate integration over an
assumed population of young pulsars. The estimates of R96 are not
applicable to very young pulsars such as the Crab, since the
synchrotron flux important for these objects is not modeled in
detail. We have chosen to increase \eff\ over the curvature value of
R96 by adding $\eta_{synch} \;=\; 10^{-3} \, (3/{\rm B}_{12}) \,
(10^3/\tau)$ to approximate this extra flux. As noted in R96 this
synchrotron flux will also contribute to gap closure, thus decreasing
the curvature \eff\ for short period pulsars.

We have computed populations over a range of parameters, but adopt
standard pulsar properties (inferred from radio studies) for this
discussion and the figures, and discuss the sensitivity of our results
to these choices below. Our standard pulsar population is born at a
rate $R = 1/100$~years at spin periods of $10$~ms, with magnetic field
distributed as a Gaussian in $\log\;$B, with mean $\log\;$B of 12.3 and
dispersion 0.3, and with magnetic inclination $\alpha$ distributed
isotropically at birth. We ignore any field evolution.

We have constructed a somewhat detailed model of the Galactic
distribution of pulsars at birth, since both local and large-scale
variations are important to the distribution of sources on the
sky. For the large-scale structure, we begin with the free electron
density model of Taylor and Cordes (1993) as a convenient tracer of
the spiral arm and inner Galaxy structure. We add to these components
a uniform density with the same radial dependence as the spiral arms:
constant within 8.5~kpc of the Galactic Center, and falling off as
$\text{sech}^2((r - 8.5~\text{kpc}) / 2 )$ beyond. The local density
variations are given by the observed distribution of O-stars. For this
purpose we created a model of the O-star population on a scale of
200~pc, smoothing the O-star density in the catalog of Garmany \et\
(1982).  We then normalized the large-scale model from the O-stars by
summing all O-stars within 0.5~kpc of a spiral arm in the approximately
complete sample within 2.5~kpc of the Sun. The resulting space density
of O-stars normalizes the arm component over the same region; O-stars
in the inter-arm region give the mean density of the uniform
component.  The local structure of the O-star distribution was then
imposed using the smoothed catalog positions; this smooth map fades
into the uniform background on a Gaussian scale length of 1.5~kpc.
The resulting distribution is shown in Figure~\ref{above}.  This gives
a reasonable picture of the large scale Galactic structure and the
local texture in the density of massive star formation.  We give
pulsars generated from this map a Gaussian scale height at birth of
$z_0~\sim~80$~pc and a Galactic $z$ velocity drawn from the 1-D
projection of the 2-D Lyne and Lorimer (1994) distribution.

\vspace{.5cm}
A surprising result is that the density model thus constructed follows
very closely on large scales the pulsar density model of Johnston
(1994), obtained from radio pulsar studies alone, increasing our
confidence as to the adequacy of our model on these scales. Locally,
the Sun is found to be in a region of unusually low density; within a
few kpc however, density is enhanced due to the Carina-Sagittarius
arm. There are also smaller enhancements in the Orion spur and Cygnus
directions, as well as at approximately 0.5~kpc in the anti-center
direction.

For completeness and in order to discuss the effects of radio
selection on our population studies, we also model the radio emission,
which originates in a physically distinct region of the magnetosphere.
We follow Johnston in modeling the radio emission and assume the
beaming evolution of Biggs (1990):
$$
    W(\text{P}) = 6.2^\circ \; \text{P}^{-1/2},
$$
and the radio luminosity law of Lorimer \et\ (1993), in standard units
of mJy~kpc$^2$ at 400~MHz:
$$
    L_{400} = 2.8\;\text{mJy}\;\text{kpc}^2\;{\dot \epsilon}^{1/2},
$$
where ${\dot \epsilon}$ is the smaller of $\text{P}{\dot \text{P}} /
10^{-15}$ and $10^2$, with a Gaussian spread in $\log L_{400}$ of 0.8.

A Monte Carlo integration of the population then begins with draws of
the pulsar properties: angle $\alpha$, age $t$, field B, $z_0$ and
$v_z$. We evolve the pulsar to its assumed age and calculate the
derived parameters $\tau$, P, ${\dot \text{P}}$, $\text{B}_{obs}$,
${\dot \text{E}}$, and $z$. We compute the correct \gam\ and radio
luminosities and beaming factors for the pulsar parameters and
determine the fraction of the sky that is swept out by the
\gam\ and radio beams to find the detection probability and
the maximum detectable distance at a given threshold.

The \gam\ threshold varies over the sky in Galactic $l$ and $b$ and is
constructed from the \EGRET\ background model and exposure map for
phases I--III, scaled so as to conform to the catalog inclusion criterion
of likelihood $\sqrt{\text{TS}}$ greater than 5 ($\sim~5\;\sigma$).
The resulting map consists of a single total photon flux sensitivity
$>$~100~MeV at each location on the sky.  The \eff\ law of R96 results in
a total energy flux from each pulsar, but not in a detailed prediction of the
pulsar spectrum, which is $\sim$~$-1.7$ depending on the pulsar age. We
assume this spectrum with a cutoff at 3~GeV in converting the photon
flux to a energy flux threshold. For comparison with other studies,
this results in a mean threshold of
$3\times~10^{-10}~\text{erg}~\text{cm}^{-2}~\text{s}^{-1}$ when
averaged over the Galactic plane.

We have not modeled the radio observations in detail, since selection
effects on P that result in large observed magnetic fields for the
young radio pulsar population are not adequately known.  The effects
of regions of massive star formation on the sensitivity of individual
surveys is also poorly known. A simple radio flux threshold of 10~mJy
approximates the average survey depth for the bulk of the young pulsar
population.

\begin{figure*}[t]
\plottwo{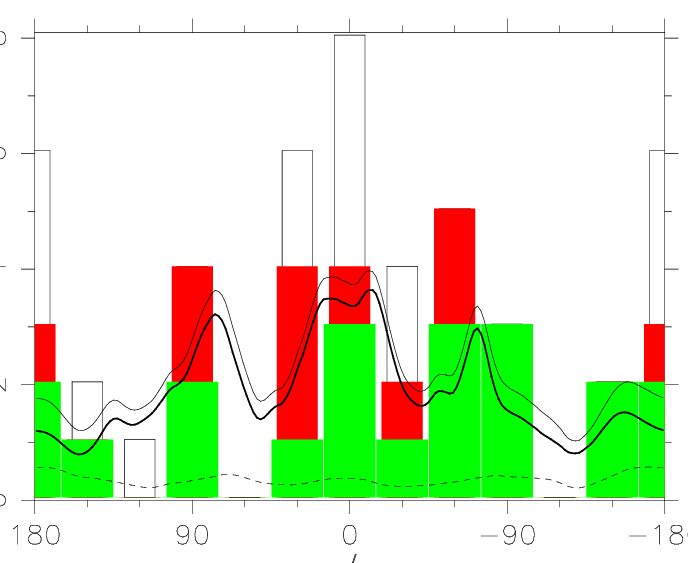}{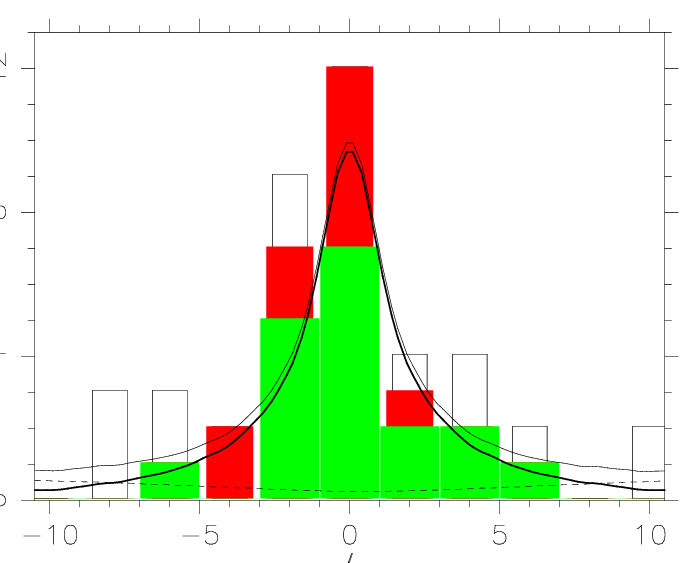}
\caption{Distribution of candidates in Galactic longitude $l$ and
latitude $b$. Histograms from darkest to lightest are for Galactic
\EGRET\ source IDs (including the known pulsars), sources with overlap
candidates, and sources with no association (outlined only). Lower dashed
curve shows the expected number of AGN seen through the Galactic disk,
upper dark curve is model prediction for the number of pulsars at a birth
rate of $1/100$~years, and the light curve is the sum of model $+$ AGN.}
\label{distrib}
\end{figure*}

\section{\label{Compare}Modeling Results}

The expected numbers of detections for the \beam\ and \eff\ laws of
R96 and our assumed pulsar population and detection thresholds are
given in Table~2 as a function of age. Most detected pulsars have ages
$\tau~\sim~10^5$~years (somewhat older than in YR due to the improved
\eff\ law of R96). As in YR only a small fraction ($<$~30~\%) of the
\gam\ objects should also be detected as radio pulsars, although a few
more might be detected in very deep searches at high DM.

As the Galaxy is transparent to \Gam, an unknown number of AGN must
be taken into account in discussing our list of \EGRET\ pulsar candidates.
From the high latitude source sample \"{O}zel and Thompson (1996) have fit
a simple power law to the AGN average flux distribution. Summing this law
over the \EGRET\ threshold map we find that a total of 4-6 sources within
$|b|~<~10^\circ$ are probably AGN.  Detailed analyses of the intermediate
and high latitude \EGRET\ sources (Grenier 1995, \"{O}zel and Thompson
1996, Nolan 1997) hint at the existence of a third class of Galactic
\EGRET\ sources with large scale height. A small number $\sim$~4 of the
$|b|~<~10^\circ$ sources could belong to this third class. A few sources
may be chance fluctuations in the background model and statistics. Since
we have found 13 high confidence IDs as well as an additional 10 overlaps,
the \EGRET\ sources are in principle accounted for.

From Table~2 our model predicts $\sim$~22 detections in the \EGRET\
range for $R~=~1/100$~years. If all 10 of the overlaps were true
associations, these together with the five known \gam\ pulsars and the
13 IDs would make 28 pulsars in the present \EGRET\ sample ---somewhat
larger than predicted by our assumed birth rate.  However, truncating
the candidate population at the 2.5~kpc ``completeness depth,'' still
leaves 19 sources, whereas we expect only 13 sources from our
model. Our counterpart catalogs are incomplete beyond 2.5~kpc so that
we cannot expect to have associated all \EGRET\ sources correctly, and
from the numbers we would then conclude that many of the overlap
candidates are chance superpositions of large nearby sources; i.e. the
true counterparts lie further from the Sun than the characteristic
catalog depth. A more likely explanation is that we have slightly
overestimated the \EGRET\ sensitivity and underestimated the local
birth rate, so that several of the nearby overlaps are in fact true
identifications.

We can compare the Galactic distributions in $l$ and $b$ of the candidate
and modeled populations. This is shown in Figure~\ref{distrib}. The model
curves match the candidate distributions successfully, both in $l$ and $b$
(at a birth rate of $1/75$~years the total number of sources, 40, is the
same as observed, and the Poisson probability of both the observed $l$
and $b$ distributions is greater than 50~\%). The density structures
associated with the spiral arms and local variations are clearly identified
in the $l$ distribution, although some of the variation, \eg\ the large
expected AGN density at $l~\sim~180^\circ$, are explained by variations
in the \EGRET\ survey sensitivity. The scale height and estimated distances
of \EGRET\ sources are consistent with our modeled population, as seen
from the $b$ distribution. As expected, no counterparts are found near
$|b|~\sim~10^\circ$. It is interesting to note that \EGRET\ sources with
no counterparts have $|b|~>~2^\circ$ and tend to lie in the direction of
the Galactic center $l~\sim~0^\circ$, hinting again at the existence of
a third class of Galactic \EGRET\ sources.

\begin{table}[t]
\addtocounter{table}{1}
\begin{center}
\caption{Expected Number of \EGRET\ Detections}
\vspace{0.5cm}
\begin{tabular}{ccccc}
$\log \tau$ range & $\gamma$ & $\gamma\;+\;$radio &
    $\gamma$ & $\gamma\;+\;$radio \\
 & & & $<$ 2.5 kpc & $<$ 2.5 kpc \\
\hline \hline
 2.0 --- 6.5 & 22.2 & 6.3 & 13.0 & 4.7 \\ \hline
 2.5 --- 3.5 &  3.4 & 0.5 &  0.5 & 0.2 \\
 3.5 --- 4.5 &  4.2 & 1.3 &  2.4 & 1.1 \\
 4.5 --- 5.5 & 11.9 & 3.7 &  7.7 & 2.9 \\
 5.5 --- 6.5 &  2.6 & 0.8 &  2.5 & 0.6
\end{tabular}
\end{center}
\end{table}

As described in \S\ref{Candidate}, we have constructed a luminosity
function for the IDs using the distances of the coincident young
objects. As we are unlikely to have associated correctly sources
beyond 2.5~kpc, we compare the model and ID luminosity functions
truncated at 2.5~kpc in Figure~\ref{luminosity}.  The K-S test
confirms that the truncated distributions are consistent.  The
histograms shown are for IDs including known sources. The luminosity
function for the known sources only, or all sources (including
overlaps) are both consistent with the luminosity function
shown. Adding the effects of radio selection is not very important
within our distance cut.

We now discuss the sensitivity of our results to modeling
uncertainties.  The mean isotropic luminosity of pulsars detected is
not sensitive to the precise details of the population model, as
pulsars of about the same flux tend to be selected at the \EGRET\
\gam\ threshold. The upper end of the untruncated model luminosity
function is dominated, however, by pulsars younger than $10^3$~years
seen at large distances, and is thus sensitive to the number of such
objects detected. As noted at the beginning of section \ref{Simulation},
very young pulsars were not modeled in detail. The spin and magnetic
history of very young pulsars is also rather poorly constrained. For
instance, doubling the initial spin period results in half as many
detections of pulsars younger than $10^3$~years. However, this has
little effect on the truncated luminosity function as very few of
these pulsars are detected within 2.5~kpc.

Varying the mean magnetic field within a reasonable range has again
little effect on the mean luminosity of the detected sources. The
total number of sources does change, increasing by 15~\% when the mean
$\log\;$B is decreased to 12.2, and conversely when increased.
Distributing the inclination angles $\alpha$ uniformly as opposed to
isotropically ($\propto \sin{\alpha}$) as suggested by Gil and Han
(1996) has essentially no effect.

\section{\label{Conclusions}Conclusions}

As reported by Kaaret and Cottam, we find that the spatial distribution
and luminosities of the low latitude \EGRET\ sources are consistent with
the proposition that most of the \EGRET\ sources are pulsars. A modest
admixure of a new class of Galactic \EGRET\ sources with large scale
height seems likely, and is hinted at in our $l$ and $b$ distributions.
Allowing for a few such sources and the extragalactic AGN seen through
the disk, we find that our outer gap model of \gam\ emission successfully
matches the number and distribution of the bulk of \EGRET\ Galactic plane
sources.

We have also confirmed the striking result of Montmerle that virtually
all of the {\em unidentified} \EGRET\ sources with a coincident SNR
are SNOBs, i.e. are also coincident with an OB association or HII
region (with the possible exceptions of G359.0-0.9 and G312.4-0.4).
We discuss this point further.

It is well known (Garmany 1994) that a large fraction of massive stars
are found outside of OB associations. The ratio of ``field'' stars to
those in associations increases with later spectral type, varying from
$\sim$~0.2 for the most massive stars, to $>$~1 for stars later than
B1, as determined from our catalog of massive stars. Runaway O-stars from
associations can realistically account for only a small fraction of
the field population, and so some isolated O-stars are formed in lower
density regions of the Galaxy. We should therefore expect to find
new-born pulsars in these same regions. In fact, if neutron stars are
still being formed from stars later than B3, we would expect to find
most pulsars outside of OB associations.

Of the known pulsars, only Vela is coincident with an OB
association. Of our 8 IDs within 2.5~kpc of the Sun, 7 are coincident
with an OB association. We thus find that 8 sources are within OB
associations, and 5 are field pulsars. In essence, this explains
Montmerle's discovery that the new \gam\ SNRs were SNOBs; in the solar
neighborhood the bright pulsars outside of OB associations had mostly
been detected as radio objects while those {\em within} OBs were still
largely unidentified.  Furthermore, many remnants outside of
associations will likely be the product of Type Ia explosions leaving
no neutron star.  If pulsars are responsible for \gam\ production,
this ensures that the sample of SNR counterparts is biased towards
those in OB associations.

We note that our present counterpart sample already provides a lower
limit on the mass of the stars which can form pulsars.  The
$\sim$~40~\% field star fraction found for the known pulsars and IDs
in the 2.5~kpc Solar neighborhood corresponds to a mean spectral class
$\sim$~B1. With a steep IMF, the cut-off mass for neutron star
formation could not then be lower than 13~$\text{M}_\odot$.  This
limit is presently still uncertain, for example assigning Poisson
errors to the number of objects allows progenitor masses as low as
8~$\text{M}_\odot$.  However, as explained in \S\ref{Compare} we
suspect that at least a few of the overlaps with OB associations are
likely to be real, so that the fraction of pulsars found in OB
associations would be even larger, and our cutoff mass higher.  In
fact, if all overlap candidates are real as suggested in KC96, then
the field fraction drops to 30~\% and the implied minimum mass for
neutron star formation is 17~$\text{M}_\odot$.  This exacerbates the
well known problem in reconciling birth rates of pulsars with massive
stars. Further refinements of this analysis, including improved
estimates for the \EGRET\ source positions and a more detailed
treatment of the displacements between birth sites and parent OBs
(\eg\ Geminga, rejected as coincident with an OB association may, in
fact, be a runaway from the Orion OB 1 association: Frisch 1993;
Smith, Cunha and Plez 1994) give promise in resolving these questions.
In particular, offsets from SNR positions due to the product neutron
star velocity would strongly support the identification of the
\gam\ sources with pulsar radiation; present positions are
however too uncertain to effect this test.

We conclude in remarking that young radio-selected pulsars are clearly
a strongly biased sample, weighted both towards long spin periods and
regions free of large dispersion and electron scattering. To some
extent, the present association of \EGRET\ sources with sites of
massive star formation selects pulsar candidates with opposite
biases (large spin-down power ${\dot \text{E}}$ corresponds to short
periods P, while proximity to massive stars results in large
$n_e$). Thus the \gam\ selected pulsar sample gives a complementary
view of the young pulsar population and a more representative census
of the recent demise of massive stars in the Solar neighborhood.

\acknowledgements

Support for this work was provided by NASA grants NAG5-3101 and
NAGW-4526.  We thank Phil Kaaret for an early description of the work
on OB counterparts, Thierry Montmerle for a detailed review of the
manuscript and Joe Fierro, John Mattox, William Tompkins and Dave
Thompson for discussion of \EGRET\ source positions.

\pagebreak


\begin{references}
\reference{} Biggs, J.D. 1990, \mnras, 245, 514
\reference{} Blaha, C. and Humphreys, R. 1989, \aj, 98, 1598
\reference{} Brand, J and Blitz, L. 1993, \aap, 275, 67
\reference{} Brazier, K.T.S. {\it et al.} 1996, \mnras, submitted
\reference{} Daugherty, J. and Harding, A.K. 1996, \apj, in press
\reference{} Esposito, J.A. \et\ 1994, \baas, 26, 970
\reference{} Fierro, J.M. 1996, Doctoral Thesis, Stanford University
\reference{} Frisch, P.C. 1993, \nat, 364, 395
\reference{} Garmany, C.D. \et\  1982, ``A Catalog of Galactic O-Type Stars,''
(http://heasarch.gsfc.nasa.gov)
\reference{} Garmany, C.D. 1994, \pasp, 106, 25
\reference{} Georgelin, Y.M. and Georgelin, Y.P 1976, \aap, 49, 57
\reference{} Gil, J.A. and Han, J.L. 1996, \apj, 458, 265
\reference{} Green D.A., 1995, ``A Catalog of Galactic Supernova Remnants
(1995 July version),'' Mullard Radio Astronomy Observatory, Cambridge, United
Kingdom
\reference{} Grenier, I.A. 1995, Adv. Space Res., 15, 73
\reference{} Halpern, J.P. and Ruderman, M. 1993, \apj, 415, 286
\reference{} Halpern, J.P. and Holt, S.S. 1992, \nat, 357, 222
\reference{} Helfand, D.J. 1994, \mnras, 267, 490
\reference{} Janes, K., Duke, C. and Lyng\aa, G. 1987, ``A Catalog of Open
Cluster Data,'' (http://heasarch.gsfc.nasa.gov)
\reference{} Johnston, S. 1994, \mnras, 268, 595
\reference{} Kaaret, P. and Cottam, J. 1996, \apj, 462, L1 (KC96)
\reference{} Kanbach, G. \et\ 1996, \aap, submitted
\reference{} Lamb, R.C. 1978, \nat, 272, 429
\reference{} Lorimer, D.R. \et\ 1993, \mnras, 263, 403
\reference{} Lyne, A.G. and Lorimer, D.R. 1994, \nat, 369, 127
\reference{} Mattox, J.R. \et\ 1996, \aap, submitted
\reference{} Merck, M. \et\ 1996, \aap, submitted
\reference{} Mel'nik, A.M. and Efremov, Yu.N. 1995,~Astron.~Let.,~21,~10
\reference{} Montmerle, T. 1979, \apj, 231, 95
\reference{} Mar\u{s}\'{a}lkov\u{a}, P. 1974, \apss, 27, 3
\reference{} Mukherjee, R. \et\ 1995, \apj, 441, L61
\reference{} Namir, E.K. \et\ 1994, \apj, 427 L95
\reference{} Nolan, P.L. 1997, in preparation
\reference{} \"{O}zel, M.E. and Thompson, D.J. 1996, \apj, 463, 105
\reference{} Romani, R.W. 1996, \apj, in press (R96)
\reference{} Romani, R.W. and Yadigaroglu, I.-A. 1995,~\apj,~438,~314~(RY)
\reference{} Smith, V.V., Cunha, K., and Plez, B. 1994,~\aap,~281,~L41
\reference{} Sturner, S.J. and Dermer, C.D. 1996, \apj, 461, 872
\reference{} Sturner, S.J. and Dermer, C.D. 1995,~\aap,~293,~L17~(SD95)
\reference{} Taylor, J.H. and Cordes, J.M. 1993, \apj, 411, 674
\reference{} Taylor, J.H., Manchester, R.N., and Lyne, A.G. 1993, \apjs, 88,
529 (http://pulsar.princeton.edu)
\reference{} Thompson, D.J., \et\ 1996b, \apjs, submitted
\reference{} Thompson, D.J., \et\ 1996, \apjs, 101, 259
\reference{} Thompson, D.J., \et\ 1994, \apj, 436, 229
\reference{} Yadigaroglu, I.-A. and Romani, R.W. 1995,~\apj,~449,~211~(YR)
\end{references}
\end{document}